\documentclass[iop,revtex4]{emulateapj}
\usepackage{graphicx}
\usepackage{times}
\usepackage{amsmath}

\begin{document}

\renewcommand{\topfraction}{1.0}
\renewcommand{\bottomfraction}{1.0}
\renewcommand{\textfraction}{0.0}
\newcommand{\eb}{\begin{equation}}
\newcommand{\ee}{\end{equation}}

\title{Chaotic rotation and evolution of asteroids and small planets in high-eccentricity orbits around white dwarfs}
\shorttitle{Chaotic rotation of high-eccentricity asteroids}

\author{Valeri~V.~Makarov}
\affil{U.S. Naval Observatory, 3450 Massachusetts Ave., Washington, DC 20392-5420, USA}
\email{valeri.makarov@gmail.com}
\author{Dimitri~Veras}
\altaffiliation{STFC Ernest Rutherford Fellow}
\affil{Centre for Exoplanets and Habitability, University of Warwick, Gibbet Hill Road, Coventry, CV4 7AL, UK}
\affil{University of Warwick, Department of Physics, Gibbet Hill Road, Coventry, CV4 7AL, UK}
\email{d.veras@warwick.ac.uk}


\begin{abstract}
Observed planetary debris in white dwarf atmospheres predominately originate from the destruction of small bodies on highly eccentric ($>0.99$) orbits. Despite their importance, these minor planets have coupled physical and orbital evolution which has remained largely unexplored. Here, we present a novel approach for estimating the influence of fast chaotic rotation on the orbital evolution of high-eccentricity triaxial asteroids, and formally characterize the propagation of their angular rotation velocities and orbital elements as random time processes. By employing the impulse approximation, we demonstrate that the violent gravitational interactions during periastron passages transfer energy between the orbit and asteroid's rotation. If the distribution of spin impulses were symmetric around zero, then the net result would be a secular decrease of the semimajor axis and a further increase of the eccentricity. We find evidence, however, that the chaotic rotation may be self-regulated in such a manner that these effects are reduced or nullified. We discover that asteroids on highly eccentric orbits can break themselves apart --- in a type of YORP-less rotational fission --- without actually entering the Roche radius, with potentially significant consequences for the distribution of debris and energy requirements for gravitational scattering in metal-polluted white dwarf planetary systems. This mechanism provides a steady stream of material impacting a white dwarf without
rapidly depleting the number of small bodies in the stellar system.
\end{abstract}


\section{Introduction}
\label{sec:intro}

Understanding the complete history and future of a planetary system usually requires investigating the host star as it changes phases. Nearly every star in the Milky Way will or already has traversed an evolutionary sequence involving main sequence, giant branch and white dwarf phases. Snapshots of white dwarf planetary systems in particular reveal exclusive and unique insights about planetary composition \citep{juryou2014,farihi2016,bonxu2017,haretal2018,holetal2018,vanrap2018,zucyou2018} and include striking examples of disintegrating \citep{vanetal2015} and intact \citep{manetal2019} orbiting minor planets.

Both major and minor planets which survive until the white dwarf phase have endured physical and orbital variations resulting from stellar evolution \citep{veras2016}. Giant branch stars induce physical variations primarily through stellar mass loss, envelope expansion and increased luminosity. Gas giant planets may accrete stellar mass through its wind, altering the composition of the planetary atmospheres \citep{spimad2012}. Envelope expansion may alter planetary surfaces and interior energy budgets through tidal interactions. Enhanced stellar luminosity may shear off atmospheres \citep{livsok1984,neltau1998,soker1998,villiv2007,wicetal2010}, melt or sublimate entire planets or some of their components, and spin up, spin down and break up minor planets through YORP-induced rotational fission \citep{veretal2014a}.

Better understood are the planetary orbital changes which are triggered by giant branch stars. Isotropic stellar mass loss expands orbits at prescribed rates \citep{omarov1962,hadjidemetriou1963,veretal2011} and anisotropic mass loss \citep{veretal2013,doskal2016a,doskal2016b} is not expected to become significant for bodies within Oort-cloud distances. Orbital variations due to stellar mass loss are effectively independent of planet mass. Tidal effects between giant stars and planets crucially set the critical displacement beyond which a planet will survive engulfment \citep{villiv2009,kunetal2011,musvil2012,adablo2013,norspi2013,valras2014,viletal2014,madetal2016,staetal2016,galetal2017,raoetal2018,steetal2018}. Effects from giant star stellar luminosity may in fact dominate over those from gravity when altering the orbits of minor planets through the Yarkovsky effect \citep{veretal2015a,veretal2019a}.

The combined result of the above forces is that as a star becomes a white dwarf, it will initially clear out the inner few au of all objects, and allow only planets larger than about 10-100 km in radius to survive intact within about 10 au. Despite these large distances, observations indicate that these objects must reach and accrete onto the white dwarf (which has a typical radius of $1R_{\oplus}$ and a typical Roche radius of $1R_{\odot}$), and do so regularly: Between one-quarter and one-half of all white dwarfs contain planetary metals \citep{zucetal2003,zucetal2010,koeetal2014}; chemical abundances of planetary debris have been measured for white dwarfs up to 8 Gyr old \citep{holetal2018}, and the white dwarf discs generated from minor planet break-up are recycled on short timescales of $~10^4 - 10^6$ yr \citep{giretal2012}.

Consequently, an important issue is how to perturb minor planets (asteroids, moons and comets) on eccentric-enough orbits ($e > 0.99$) to reach the white dwarf. Many methods have been successfully invoked: gravitational perturbations by a single planet \citep{bonetal2011,debetal2012,frehan2014}, multiple planets \citep{veretal2016,payetal2016,payetal2017,musetal2018,smaetal2018}, multiple stars \citep{bonver2015,hampor2016,petmun2017,steetal2017,steetal2018} and Galactic tides \citep{alcetal1986,paralc1998,veretal2014b,stoetal2015,caihey2017}. The outcome of a minor planet entering the Roche radius of a white dwarf has also been explored \citep{debetal2012,veretal2014c,wyaetal2014,veretal2015b,broetal2017}, particularly with regard to the resultant disc evolution \citep{bocraf2011,rafikov2011a,rafikov2011b,metetal2012,rafgar2012,kenbro2017a,kenbro2017b,mirraf2018}. Star-planet tides negligibly affect the orbital motion of minor planets \citep{veretal2019b}, such as the ones uncovered by \citet{vanetal2015} and \citet{manetal2019}.

However, in nearly every case listed above, the physical evolution of asteroids on highly eccentric orbits was not modelled, nor was the resulting feedback on the orbital angular momentum. This paper strives to address this missing component of our understanding. In Section 2 we describe our procedure for simulating the coupled evolution of these asteroids. In Section 3 we argue that a highly-eccentric asteroid need not encounter the Roche radius before spinning itself apart. Section 4 contains a formal analysis of our simulation output in terms of random processes which describe the evolution of the physical and orbital elements. We derive functional dependencies of the orbital evolution on these random processes in Section 5 and discuss and summarize our results in Section 6. We henceforth refer to our orbiting objects as ``asteroids''.

\section{Chaotic rotation}
\label{cha.sec}
Rotation of elongated bodies around their principal axes of inertia in two-body systems are mostly driven by the gradient
of the force of gravitational attraction exerted by the orbiting companion. The tidal interactions are much smaller
in magnitude for non-vanishing parameters of triaxiality $\sigma=(B-A)/C$, which are often found in smaller planets and
asteroids. The $A$, $B$, and $C$ here and in the following designate the principal moments of inertia in increasing order.
In this paper, we use the term ``asteroid" in its generic meaning, which includes celestial bodies rigid enough to maintain a
permanent shape, such as comets, asteroids, and minor planets. Our results are applicable to a wide range of such objects,
as long as they have a significantly prolate shape and highly eccentric orbits.
Rotation of triaxial asteroids is known to have complex, structured sections in the parameter space \citep{wis}. The
islands of stable equilibrium in the Poincar\'e sections are surrounded by bands of purely chaotic motion, which are present
even for small eccentricity orbits as long as $\sigma$ is finite. In the narrow boundary zone between chaos and libration zones,
small secondary resonance islands may be present \citep{fly}, but their physical validity remains to be confirmed. Only 
Hyperion and, possibly, Prometheus and Pandora, represent satellites in chaotic rotation in the Solar system \citep{kou, mel}.
But chaotic satellites are likely common around minor planets in the outer Solar System. Many TNOs have eccentric satellites, and a 
particularly instructive example is the satellite Thorondor around the large TNO (385446) Manw{\"e} \citep{gru}, which has an eccentricity of 0.56.
The prevalence of synchronous rotation can probably be explained by the regularization effect of tidal dissipation in
dynamic systems of small and moderate eccentricity. The synchronously rotating satellites of significantly elongated shape,
such as Phobos and Epimetheus, could not enter their current spin-orbit state without having to cross the chaotic zone
\citep{wis87}.

The case of high eccentricity ($e>0.95$) and significant triaxiality is different, because chaotic behavior is the only possible
state for such asteroids. The islands of stable equilibrium vanish, and the tidal dissipation can not regularize the motion
even in the long run. We investigate this case by means of numerical simulations for the simplest setup, in which the orbit and
the asteroid's equator are coplanar (i.e., the obliquity of the orbit on the equator is zero) and the problem becomes one-dimensional.
The tumbling rotation of prolate asteroids is certainly three-dimensional in reality, and this approximation is
likely to slightly overestimate the associated acceleration. 
Tidal forces are ignored. Our model object of choice is similar to Proteus in size, mass, and shape, but we put it into a long-period,
high-eccentricity orbit around a white dwarf primary (see Table~\ref{param.tab} for our adopted physical and orbital elements). The semimajor axis of 1.5 au was chosen for computational feasibility (see Section 4), and we expect the results to be qualitatively similar for higher values of $a$. The orbital parameters are constant, an assumption which is justified for a relatively short integration spanning less than $10\,000$ orbits in view of the relatively slow rate of orbital evolution, as described below.

 \begin{table*}
 \centering
 \caption{Model parameters used in simulations. The semimajor axis corresponds to 1.5 au,
 the white dwarf mass to $0.6 M_{\odot}$ and the asteroid mass to a value which is about 5\%
 the mass of Ceres.}
 \label{param.tab}
 \begin{tabular}{@{}lrrr@{}}
 \hline
   Name     &  Description    & Units & Value\\
 \hline
 $a$ & \dotfill semimajor axis & m & $2.244\cdot 10^{11}$\\
 $e$ & \dotfill orbital eccentricity & & variable ($>0.95$)\\
 $M_1$ & \dotfill mass of primary body(white dwarf)& kg & $1.193\cdot 10^{30}$\\
 $M_2$ & \dotfill mass of asteroid & kg & $4.4\cdot 10^{19}$\\
 $\sigma=(B-A)/C$ & \dotfill triaxiality of asteroid &  & $0.05$ \\
 $R$   & \dotfill radius of asteroid  & m & $2.1\cdot 10^5$\\
 $\xi$   & \dotfill coefficient of inertia  &  & $0.35$\\
 \hline
 \end{tabular}
 \end{table*}

The integrated ordinary differential equation (ODE) is
\eb
\theta''(t)+\frac{3}{2}n^2 \frac{\sigma\;\sin(2\,\theta(t)-2\,\nu(t))}{\left(1-e\,\cos E(t)\right)^3}=0,
\label{ode.eq}
\ee
where $t$ is time, $n$ is the mean orbital motion, $\nu(t)$ is the true anomaly in radians, $E(t)$ is the eccentric anomaly in radians,
and $\theta$ denotes the orientation angle of the asteroid counted from its longest axis in the inertial space \citep{danb}. The functions of the time-variable
anomalies $\nu$ and $E$ are often replaced with well-known series in powers of eccentricity, relating them to the linearly changing mean anomaly.
The main challenge of this problem is that such series become impractical at large $e$ because an exponentially growing number of terms has to
be used, and the associated Kaula's functions of eccentricity become numerically large. We have to revert to the traditional computation of
$E$ using Kepler's equation. We made use of the technique of inverse interpolation proposed by \citet{tom} to compute $E$ as a function of mean anomaly.
We also employed an integration scheme with self-adaptive time steps to capture the all-important sharp variations of $\theta$ that take place during
the periastron passage. The time step has to be much shorter than the typical duration of this passage, while a coarser time step is
prudent to use outside this orbit phase interval, i.e., for 99.9\% of the orbit. The ODE (\ref{ode.eq}) can be integrated with two additional
boundary conditions for $\theta(0)$ and $\theta'(0)$.

\begin{figure}
\centerline{
\includegraphics[width=8.5cm]{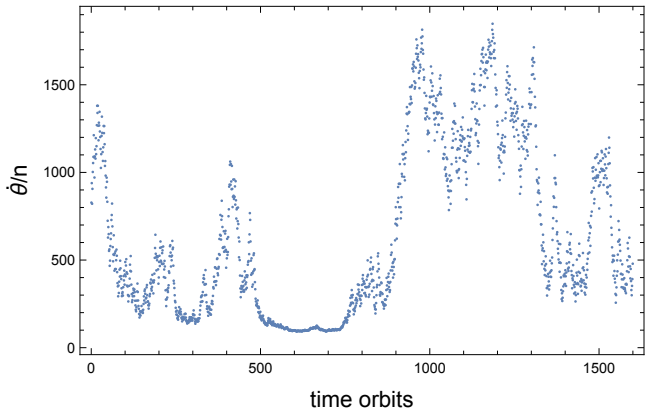}
}
\caption{Simulated evolution of apoastron rotation velocity of a Proteus-like asteroid orbiting a white dwarf on a
highly eccentric ($e=0.99$), long-period ($a=1.5$ AU) orbit.
\label{rot.fig} }
\end{figure}

Fig. \ref{rot.fig} shows a typical result of numerical simulation of 1700 orbits for $e=0.99$. It shows the normalized apoastron rotation velocity
$\dot\theta/n \equiv \omega/n$ versus orbit number, for initial conditions $\theta(0)=-0.04$ rad, $\theta'(0)=801\;n$. The behavior is chaotic with a short Lyapunov
exponential; hence, even a marginally small perturbation of the initial values brings about a completely different rotation curve. However, important
observations can be made based on this limited example. The spin rate varies in a very broad range, but it seems to be confined to a certain interval
of prograde rotation. It bounces back when it comes close to zero. The asteroid's rotation is fast most of
the time, but can it become arbitrarily fast? The first idea is that the velocity may be limited to the maximum orbital motion at the moment
of closest approach, which is computed as 
\eb
\dot\nu_{\rm max}/n=\frac{\sqrt{1+e}}{(1-e)^\frac{3}{2}}.
\label{nu.eq} 
\ee
A few characteristic values are 124.9 at $e=0.95$,
1410.7 at $e=0.99$, 15803.5 at $e=0.998$. The results in Fig. \ref{rot.fig} for $e=0.99$ indicate that the spin rate can become higher than
$\dot\nu_{\rm max}$ even on a relatively short time scale. Indeed, integrations with other initial conditions revealed that the rate of rotation may
become as high as $3000\,n$. We give more consideration to the issue of the highest rate of rotation in \S~\ref{rp.sec}.

\section{Rotational fission versus tidal break-up}

Much attention has been paid in the literature to the issue of rotational fissure of Solar system asteroids \citep{wal, ort, vok},
which may account for a significant population of binary asteroids of small mass ratio. Observations show that most
minor planets with diameters smaller than 10 km have a lower bound of $\sim2$ hr for their rotation periods \citep{pol}. Within the
framework of a ``rubble pile" model, where the bodies are held together predominantly by self-gravitation, the critical
spin rate of fission depends on the diameter and other, less observable parameters. Larger asteroids with diameters greater
than 10 km rotate slower than $\sim4$ hr \citep{war}. It is therefore of interest to know if impulse-like excitation of rotation
at perihelion can drive asteroids to critical spin rates, and how large the eccentricity should be ($\equiv e_{\rm critical}$) for that to happen.

\begin{figure}
\centerline{
\includegraphics[width=8.5cm]{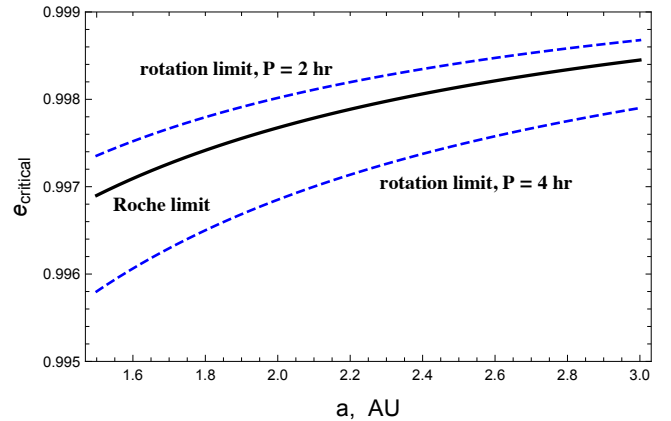}
}
\caption{Critical values of orbital eccentricity for rotational fission by the centrifugal force
  (dashed lines) or by the tidal force within the Roche radius (solid curve). The vertical order
  of the curves is independent of semimajor axis.
\label{ecrit.fig} }
\end{figure}

Fig. \ref{ecrit.fig} shows the values of critical eccentricity computed with these simplifying assumptions: 1) tidal fission
takes place when the periastron distance $a\,(1-e)$ becomes smaller than $1\,R_{\sun}$, which is approximately the Roche radius
for white dwarfs; 2) the highest rate of rotation is close to the maximum periastron angular velocity, Eq.~\ref{nu.eq}. Rotational
fission of larger asteroids, such as our Proteus analog, is achieved at lower eccentricity than the tidal break-up. The difference
in eccentricity may seem small, but the difference in the periastron distance is significant. This result suggests
that smaller parts and debris can agglomerate around white dwarfs well outside the Roche radius, to be delivered to the
stellar surface by other means.

We already noted in \S~\ref{cha.sec} that the actual spin rate can become much higher than $\dot\nu_{\rm max}$, as follows from our
numerical simulations. A single impulse that drove the asteroid to spin-up above the critical
value is enough to break it. Therefore, fission
by centrifugal force is likely to be more efficient, especially for larger asteroids, because it can happen at a lower
eccentricity. More discussion of the upper boundary of rotational velocity is given in \S~\ref{rp.sec}.

\section{Rotation velocity as a random process}
\label{rp.sec}
\begin{figure*}
\centerline{
\includegraphics[width=8.5cm]{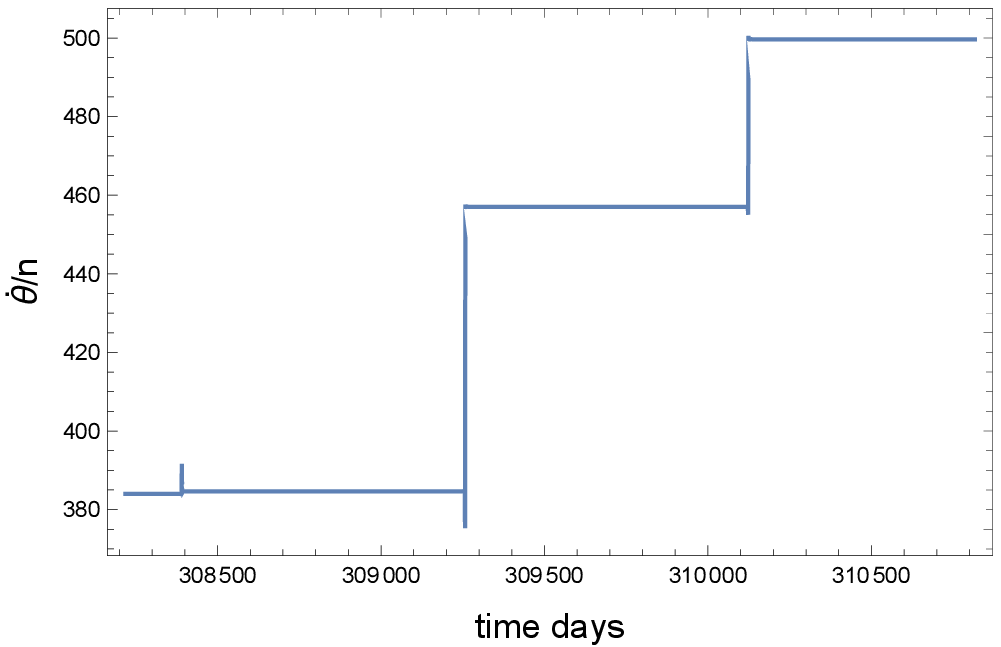}\includegraphics[width=8.5cm]{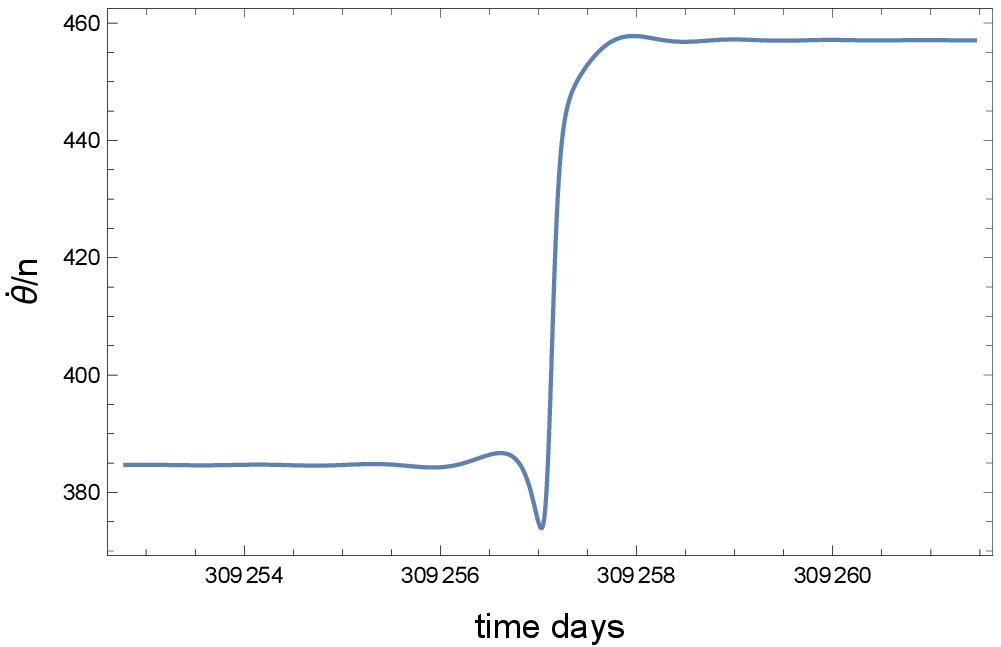}
}
\caption{A segment of the simulated rotation velocity curve for the model described in Table~1 spanning three orbits (left),
and the same curve shown in more detail for one periastron passage (right).  
\label{3orb.fig} }
\end{figure*}

Figure~\ref{rot.fig} shows a rotation velocity curve obtained by numerical integration. The large changes of velocity
without any pattern or periodicity suggest that this may be a random process. This figure shows only the values
of velocity computed at apoastrons where the triaxial torque is at a minimum (Eq.~\ref{ode.eq}). Fig.~\ref{3orb.fig}, left panel, depicts how
the simulated velocity changes within each orbit. For most of the time, the variation is relatively slow, but during a
periastron encounter, the acceleration increases by a factor $\sim1$ million (at $e=0.99$), leading to an almost
step-wise change in velocity (Fig.~\ref{3orb.fig}, right panel). In view of these findings, the impulse approximation is the
most suitable  model for long-period asteroids with $e>0.95$.

We assume in the impulse approximation that all the dynamical interaction takes place momentarily during the closest
approach to the star at the periastron. The result of this interaction is an abrupt change in rotation velocity $\omega\equiv\dot\theta$.
It is sufficient then to measure the velocity at the apoastron where $\nu=E=\pi$, and to consider the difference between two consecutive apoastron velocities $d_{\omega,i} = \omega_{i+1}-\omega_{i}$ as the impulse magnitude. The discrete sets
of tuples $\{t_i,\omega_i\}$ and $\{t_i, d_{\omega,i}\}$ are considered to be random processes in time.

We performed several numerical integrations of ODE~\ref{ode.eq} with varying initial conditions for fast prograde rotation velocities,
$e=0.99$, and 9000 orbits. The solutions were sampled at apoastron times and analyzed as random processes using the Wolfram Mathematica
{\tt TimeSeriesModelFit} function\footnote{\url{https://reference.wolfram.com/language/ref/TimeSeriesModelFit.html}}. 
This function determines which kind of random process best describes the given sequence of data and
fits the corresponding model parameters. For the normalized rotational velocity $\omega/n$, the most frequent output is
an {\it Autoregressive Moving Average} process ARMA$(1,2)$. A time series of values $z_t$ can be represented in this model
as
\eb
z_i=c+\phi\,z_{i-1}+\epsilon_i+\psi_1\,\epsilon_{i-1}+\psi_2\,\epsilon_{i-2},
\ee
where $c$ is a constant, $\epsilon_i$ is a sequence of independent random numbers with a zero mean and a variance $\varpi^2$, and $c$, 
$\phi$, $\psi_1$, and $\psi_2$
are the fitting model parameters. The presence of a nonzero $\phi$ demonstrates the auto-regressive part of the model, in that
each outcome depends on the previous value. Generally, the process is stationary if $|\phi|<1$. The one or two nonzero parameters $\xi$ indicate
that the random impulses are correlated with one or two preceding impulses, which represents the moving average property. It may
seem puzzling why the impulse magnitude, which depends only on the apoastron velocity and orientation angle, is correlated
with the preceding impulses, but the reason will be revealed in the following analysis of $d_\omega$ properties. 

One particular integration and model fitting for 9000 orbits produced these results: $c=4.5$, $\varpi^2=3010$, $\phi=0.998$, $\psi_1=0.009$,
$\psi_2=-0.034$. Although the values vary between simulations, some common features are present. The $\phi$ value is close to unity,
which means the process is weakly stationary. Starting the process with ``unreasonable" initial conditions results in an extended
transitional phase before the behavior settles down and stabilizes, which explains why long integrations covering many orbits are
required to produce a reliable output.  Because the root of the
autoregressive polynomial is greater than 1, the process is causal. The values of $\psi_1$ and $\psi_2$ are small, thus,
the impulses are weakly correlated. One of
the roots of the moving average polynomial is negative, so the latter is not invertible. Finally, the positive value of $c$ indicates
that the rotation velocity has a tendency to infinitely grow. However, this result is doubtful in view of complications, which we
will now discuss.

Generally, we do not find the fitted ARMA(1,2) process to adequately represent the simulated velocity curve. Some important features
seem to be missing in this model. For example, nothing prevents an ARMA(1,2) process with the estimated parameters to descend to
negative values, whereas our simulations are always confined to a range of prograde rotation. Although a stationary process is
intuitively expected, the moving average part appears to be a distorted reflection of more intricate properties of the system,
which are not captured by this simple model. We believe the main reason for these shortcomings is the peculiar distribution
of velocity impulses, which is neither Gaussian, nor identical for the process instances.

More progress can be made if we consider the time series of velocity impulses $d_\omega$ as a random process. The model fitting
procedure invariably produced a {\it Generalized Autoregressive Conditionally Heteroscedastic} model GARCH(1,1). A discrete-time
GARCH(1,1) process $x(t)$ is a series of statistically independent random numbers with a zero conditional mean and a conditional variance
\eb
\mathcal{E}[x_i^2\,|\,\{x_{i-1}\}]=\varpi^2_i=\kappa+\alpha_1\, x_{i-1}^2 +\beta_1\,\varpi^2_{i-1},
\ee
where $\kappa$, $\alpha_1$, and $\beta_1$ are positive-definite model parameters. This fit captures one additional curious
property of the process under investigation, viz., the variance of a velocity pulse depends on both the variance and the realization
of the previous pulse. The unconditional variance of $x(t)$ is $\kappa/(1-\alpha_1-\beta_1)$, so that for a weakly stationary process
$\alpha_1$ and $\beta_1$ should be between 0 and 1. 

A particular numerical integration produced these model fit parameters for $d_\omega/n$: $\kappa=28.2$, $\alpha_1=0.119$, $\beta_1=0.872$, which represent a typical outcome of this simulation.
We note the strong dependence of the current variance on the preceding variance. There is no requirement that the underlying
distribution is Gaussian. If the process wanders into a domain of small impulse values, the dispersion of impulse  values is likely
to remain small for a period of time. In other words, the velocity is likely to be less variable in the future if the previous
variations have been small. This is similar to the behavior of the financial market price volatility, for example, where extended
periods of lull are separated by periods of elevated variability.

\begin{figure*}
\centerline{
\includegraphics[width=8.5cm]{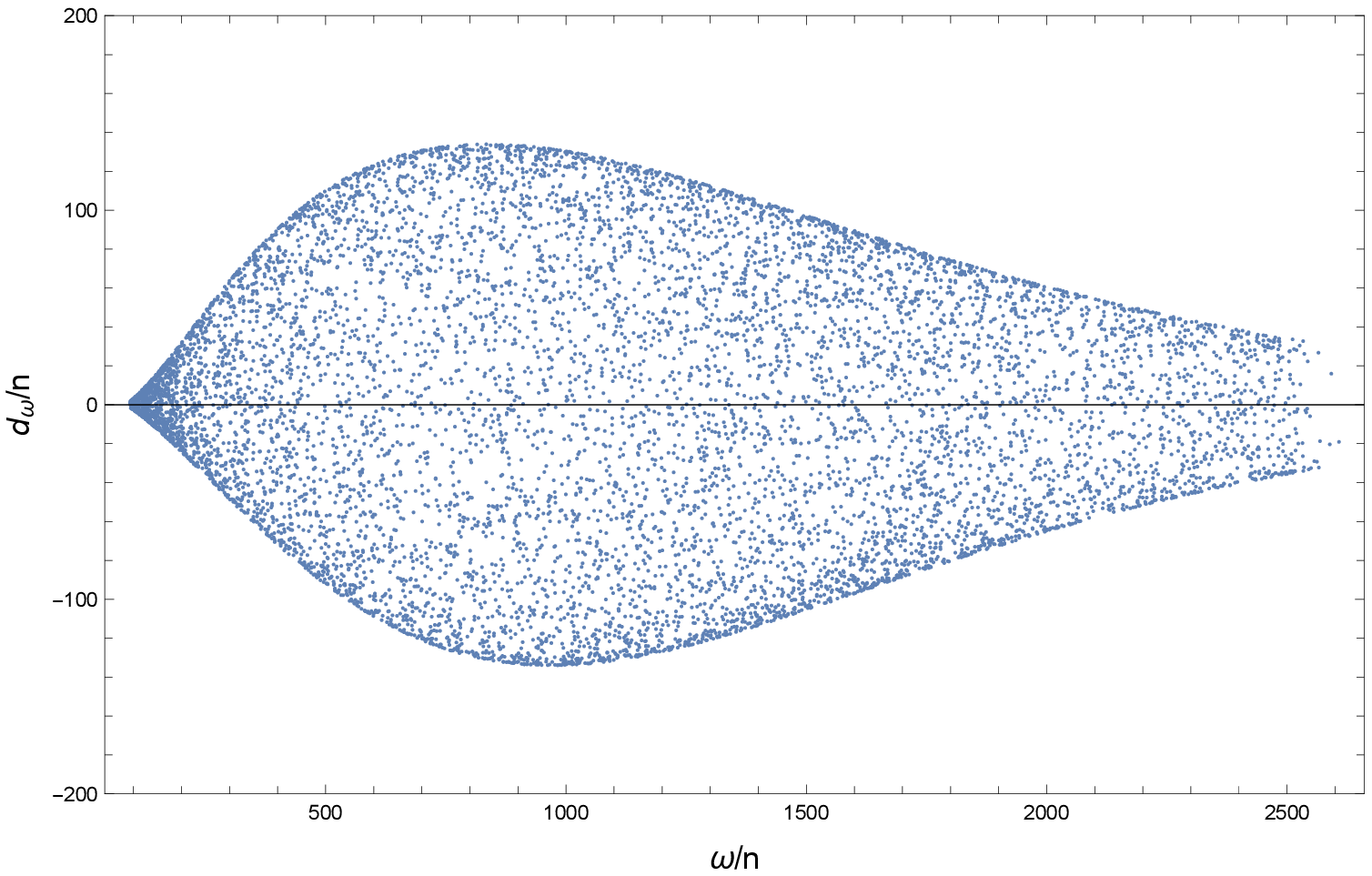}\includegraphics[width=8.5cm]{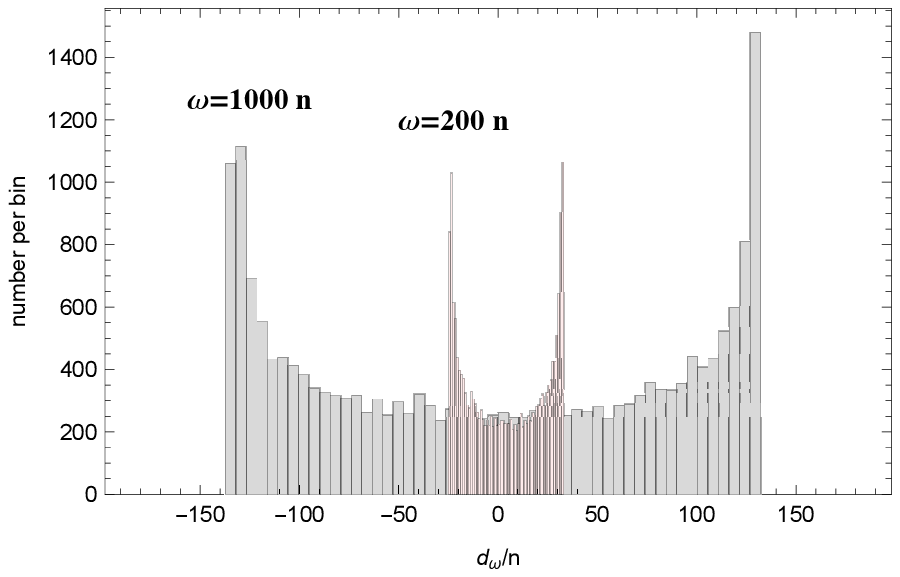}
}
\caption{Left: Normalized perturbations of spin rate versus apoastron values of spin rate computed in a numerical
simulation of 9000 orbits for $e=0.99$ and star-planet parameters from Table~1. Right: Histograms of $d_\omega$ perturbation
distributions at fixed apoastron velocities $\omega=200\,n$ and $1000\,n$ computed by Monte-Carlo simulations
of $20\,000$ periastron passages with random initial orientation angles.
\label{map.fig} }
\end{figure*}

The origin of these peculiarities becomes more understandable when we map the apoastron velocity pulses versus velocity values,
analogous to the Poincar{\'e} section mapping. Fig.~\ref{map.fig}, left, shows a realization obtained by numerical integration
over 9000 orbits with $e=0.99$, $\theta(0)=-0.02$, $\dot\theta(0)=800\, n$, and other parameters from Table~1. The map confirms
that the velocity is limited by $\sim100\,n$ at the minimum, but there seems to be no upper bound. More surprisingly,
the domain of possible rotation rates and impulses is confined in the latter on both sides. For any specific apoastron
velocity, the distribution of periastron impulses is finite. The chaotic nature of the process is betrayed by the
random position of states within the domain. When the velocity stochastically reaches the smallest possible value,
the process is ``cornered" in such a way that only vanishingly small velocity updates are possible, and the process
lingers in this low-volatility regime for an extended period of time. The distribution of velocity impulses is not
symmetric around zero. The largest positive (prograde) updates happen at a lower velocity than the largest
negative (retrograde) updates. Therefore, when the velocity is smaller than $\sim 800\,n$, greater prograde impulses
are possible and the velocity is likely to wander towards faster values. The situation reverses when it becomes
greater than $\sim 1000\,n$, where retrograde impulses of greater magnitude become possible, which will likely succeed
in pushing the velocity back. This peculiar distribution allows the process to remain stationary and in the long run,
vary around some median velocity range.

We performed targeted numerical simulations of the periastron passage (instead of a continuous integration for 9000 orbits)
to find out the properties of the $d_\omega$ distribution. For a fixed input velocity $\omega$, $20\,000$ Monte-Carlo
integrations are performed with random input orientation angle $\theta$ covering a short interval around the periastron
time. The resulting sampled distributions for $\omega=200\,n$ and $1000\,n$ are shown in Fig.~\ref{map.fig}, right.
The width of the distributions strongly depends on the input velocity. There is also a pileup at both ends of the
interval of allowable perturbations, so that impulses of greatest magnitude are more frequent than weak impulses.

These results explain why an ARMA model does not capture the properties of the $\omega(t)$ process well enough --
the distribution of ``noise" is drastically non-Gaussian. The GARCH model may be more successful in representing
the $d_\omega$ process, because it does not require the distribution to be Gaussian. However, the variance as
a model parameter is perhaps of limited use for the observed concave distribution of velocity updates.

\section{Orbital evolution of high-eccentricity asteroids}

Each periastron passage of a high-eccentricity asteroid results in an abrupt change of its rotation velocity.
If we ignore the dissipative tidal forces in the asteroid and the star, the total energy of the closed system
should be constant:
\eb
E_{\rm orb}+E_{\rm rot}= -G\frac{m_1 m_2}{2a}+\frac{\xi m_2}{2}R^2\omega^2=const.
\ee
A change in $\omega$ results in a change of the orbit semimajor axis $a$. Taking into account that the asteroid's
radius $R<<a$ and its mass $m_2<<m_1$, the change in $a$ is readily derived as
\eb
\frac{d_a}{a}\simeq -\frac{\xi (2\omega\,d_\omega+d_\omega^2)}{n^2}\left(\frac{R}{a}\right)^2.
\ee
The quadratic term $d_\omega^2$ may indicate that the orbit should be secularly shrinking. However, this is only true
if the distribution of velocity updates $d_\omega$ were symmetric around zero, which is not the case. As Fig.~\ref{map.fig}
(left) indicates, the distribution may be biased toward negative values at high rotation velocities, compensating for
the positive secular term or even causing the orbit to expand. Our limited numerical experiments are nor sufficient
to verify this. It is safe to conclude that the orbit updates also constitute a random time process. The fractional
updates of the orbit are found to be small for the model parameters in Table 1 (of the order of $10^{-9}$), but they
can be much larger for super-earth exoplanets.

The total angular momentum should also be conserved in this two-body closed system:
\eb
\begin{split}
L_{\rm tot}=L_{\rm orb}+L_{\rm rot}=G^\frac{1}{2}\frac{m_1 m_2}{(m_1+m_2)^\frac{1}{2}}\sqrt{a(1-e^2)}\\
+\xi\,m_2\,R^2\omega=const.
\end{split}
\ee
Updating the $a$, $e$, and $\omega$ by their increments, equating the change in the angular momentum to zero, and
taking into account that $m_2<<m_1$, $d_e<<1$, $d_a<<1$, the following approximate equation obtains
\eb
\frac{d_e}{e}\simeq  \frac{\xi \sqrt{1-e^2}}{n\,e^2}\left(\frac{R}{a}\right)^2\,d_\omega.
\ee
The fractional update of eccentricity $d_e/e$ is approximately linear in the update of rotation velocity
$d_\omega$. The outcome of this stochastic variation is, again, uncertain, because it depends on the
distribution of $d_\omega$. Because of the asymmetry observed in Fig.~\ref{map.fig}, left, the eccentricity
is expected to grow when the asteroid rotates relatively slowly, but is likely to decline when the asteroid
rapidly spins. The overall long-term evolution is unclear from our limited numerical integration.

\section{Discussion and conclusions}
\label{sec:disc}
Exchange of angular momentum and energy between the orbit and the spin during close periastron passages of high-eccentricity
asteroids causes the angular rotation velocity and the orbital elements to chaotically change. These processes can be
formally modeled as random time processes. The long-term outcome for the orbit is unclear from our limited integration
runs, but the indication is that the processes are weakly stationary. Any consideration of the orbital evolution is
incomplete without the dissipative tidal force: the tides in the asteroid are probably negligible because of the explicit
dependence of orbital action on $(R/a)^5$ \citep{mak}. Observational statistics for tumbling Solar system asteroids
suggest that tidal dissipation inside asteroids provides long-term damping of the tumbling (chaotic, three-dimensional)
motion, especially for high tensile rigidity objects \citep{pra}. However, this effect may be greatly diminished by the inverse
proportionality of the tidal quality function to tidal frequency \citep{efr} for rapidly rotating high-eccentricity asteroids
around white dwarfs. The tides in the star may be more efficient on the timescale of
stellar ages despite their expected low values of the tidal quality factor. Most stars rotate slower than the maximum angular velocity
of the asteroid at the periastron, where practically all the action takes place. The tidal bulge raised on the star lags
the direction to the asteroid, and the resulting torque gradually circularizes and shrinks the orbit. 

We chose Proteus as the model object for our simulations (Table~1), which is larger, more massive, and more spherically symmetric
than a typical Solar system asteroid, but the results are relevant to a broad range of objects from small comets to minor planets.
The equation of motion (\ref{ode.eq}) is independent of mass, but the orbital evolution is sensitive to the secondary mass. It is
of interest to probe much greater triaxiality values, such as Hyperion at $\sigma\simeq 0.1$ and super-Earth exoplanets with
low $\sigma$ ($\sim 10^{-5}$). Our preliminary computation for a Hyperion analog indicate that the behavior described in
this paper and represented by Figs. \ref{rot.fig}, \ref{3orb.fig}, and \ref{map.fig} is also found in smaller and more prolate
bodies, but the velocity impulses are proportionally greater and their distribution around zero is more asymmetric. Such objects are
likely to stochastically achieve break-up spin rates faster than the larger bodies. Conversely, they may become a viable source
of white dwarf pollution at lower orbital eccentricity.

Although the behavior of rotation velocity is obviously chaotic and stochastic, modeling it with standard types of random
processes is only partially successful. The reason for this is the peculiar distribution of velocity impulses, which is
finite at any given velocity above the minimum prograde value. The probability of a velocity impulse sharply increases
toward the boundaries of the distribution, which are not symmetric around zero. As a result, the process is weakly
stationary and somewhat self-regulating. If the spin rate becomes low, the range of allowable impulses tends to zero,
and the asteroid may linger in this quiescent state for an extended period of time. The high-velocity end of the distribution,
on the other hand, does not seem to be bounded. The spin rates may become very high, but the probability of that is small,
and it may take a long time for the asteroid to wander into this regime.

After the asteroid reaches a particular spin threshold, it will be disrupted by the centrifugal force. Our calculations illustrate that rotational fission may be more efficient than tidal disruption, especially for the largest asteroids and minor planets, because fission is triggered at lower values of eccentricity. The consequences for polluted white dwarfs are two-fold: (1) debris produced from break-up would populate a radial region which extends beyond the Roche radius, and (2) the orbital eccentricity which metal-polluting asteroids need to acheive is slightly lower than previously thought. We caution that we are unable to obtain a likelihood of rotational fission (versus tidal disruption) nor predict a specific asteroid's long-term evolution, due to both computational limitations and the stochasticity of the systems. Nevertheless, this study represents a first step in the exploration of the coupled spin and orbital evolution of the dominant source of the observational signatures of the vast majority of known white dwarf planetary systems.

\section*{Acknowledgments}

DV gratefully acknowledges the support of the STFC via an Ernest Rutherford Fellowship (grant ST/P003850/1).

\end{document}